\documentclass[10pt,aps,amsmath,showpacs,amssymb,nofootinbib,pre,twocolumn,epsfig]{revtex4}
\usepackage{bmpsize}
\usepackage[dvips,usenames]{color}
\usepackage{amsmath}
\newcommand{\av}[1]{\langle {#1} \rangle}

\usepackage{graphicx}
\usepackage{lipsum}

\begin{document}

\title{Non-universal power-law dynamics of SIR models on hierarchical 
modular networks} 

\author{G\'eza \'Odor}
\affiliation{Institute of Technical Physics and Materials Science,
Center for Energy Research, P. O. Box 49, H-1525 Budapest, Hungary}


\date{\today}


\begin{abstract}


Power-law (PL) time dependent infection growth has been reported in many 
COVID-19 statistics. In simple SIR models the number of infections grows at the
outbreak as $I(t) \propto t^{d-1}$ on $d$-dimensional Euclidean lattices 
in the endemic phase or follow a slower universal PL at the critical point, 
until finite sizes cause immunity and a crossover to an exponential decay.
Heterogeneity may alter the dynamics of spreading models, spatially 
inhomogeneous infection rates can cause slower decays, posing a threat 
of a long recovery from a pandemic. 
COVID-19 statistics have also provided epidemic size distributions with PL
tails in several countries.
Here I investigate SIR like models on hierarchical modular networks,
embedded in 2d lattices with the addition of long-range links. 
I show that if the topological dimension of the network is finite, 
average degree dependent PL growth of prevalence emerges. 
Supercritically the same exponents as of regular graphs occurs, 
but the topological disorder alters the critical behavior.
This is also true for the epidemic size distributions.
Mobility of individuals does not affect the form of the scaling 
behavior, except for the $d=2$ lattice, but increases the magnitude 
of the epidemic. The addition of a super-spreader hot-spot also does 
not change the growth exponent and the exponential decay in the herd 
immunity regime.
\end{abstract}

\maketitle


\section{Introduction}


Human infectious diseases usually start with an exponential growth, 
as the number of healthy neighbors is high, thanks to the small world 
connectedness of societies. Thus a full, infinite dimensional
graph approximation, described by mean-field behavior is valid.
In finite $d$ dimensions this evolution is slower and in case of 
the Susceptible Infected Recovered (SIR) process~\cite{PastorSatorras2015}, 
the simplest model for epidemics with immunization, the actual number of
infected individuals follows a scaling behavior to leading order:
$I(t) \propto t^{d-1}$.
However, this is true in the supercritical phase, where the reproduction 
number of epidemiology is $R_0 > 1$. By reducing $R_0$ a continuous 
phase transition to a non-endemic state happens and right at the critical 
point: $R_0 = 1$ the number of infected individuals grows algebraically, 
characterized by the so-called initial slip exponent $\eta$ in the scaling
law $I(t) \propto t^{\eta}$ of statistical physics~\cite{marro2005,HHL,odorbook}.
For different models the value of $\eta$ is known in different Euclidean 
dimensions of the substrate graphs~\cite{Munoz99,odorbook}, in 
homogeneous systems. 

In the case of quasi-static, or quenched heterogeneity much less is known.
According to the Harris criterion~\cite{Harris}, the disorder is 
irrelevant for the critical behavior of SIR, belonging to the
Dynamical Isotropic Percolation (DIP) universality class~\cite{gras83,odorbook}.
This means that the correlation length exponent $\nu_{\perp}$ 
fulfills the inequality: $d \nu_{\perp} > 2$ and weak disorder 
decreases under coarse gaining becoming unimportant on large 
length scales. 
But this criterion is a necessary and not sufficient condition 
for the stability of the impure fixed point and has been found to
fail in certain models of nonequilibrium statistical 
physics~\cite{Vojta2006b}.

Containment measures can push $R_0$ below 1 by lowering the
infection rate or the graph dimensions for example.
Large amount of Covid data have provided various growths functions
of $I(t)$ with PL, exponential or mixed time 
dependence~\cite{doi:10.1098/rsif.2020.0518} in different
countries and at different times.
 
Very recently a basic SIR like model has been investigated on Euclidean
lattices with inhomogeneous infection rates, including the possibility
of mobility~\cite{sakaguchi2020slow}. 
A striking numerical conclusion was drawn, that in the presence of 
super-spreader hot spots, where the infection rate is much higher than 
the average, the epidemic does not vanish exponentially fast by herd 
immunity, but decays in a slow PL manner way. This behavior was paralleled 
with the Griffihts Phase (GP) phenomena~\cite{Vojta2006b}, which occurs near
the critical points of phase transitions in strongly heterogeneous
systems. At first glance this seems surprising, because for having GP
one wold need long surviving rare regions (RR), in which the 
activity disappears exponentially slowly by the region size: 
$\tau_t \sim e^{V}$, while in the SIR model recovered individuals cannot 
be re-activated but become inactive forever. 
Thus the effective topological dimension of an infected region
decreases quickly as herd immunity develops.
The authors of Ref.~\cite{sakaguchi2020slow} suggest a clue for this 
strange behavior by the interplay of SIR processes and diffusion. 
Indeed, mobility can increase the effective dimension of systems 
and in the infinite diffusion limit mean-field behavior emerges, 
with a PL decay at criticality.
Furthermore, mobility of individuals can transform recovered sites susceptible
again and Susceptible Infected Susceptible (SIS) type of scaling behavior, 
belonging to the Directed Percolation (DP)~\cite{marro2005,HHL,odorbook} 
universality class may be observed.

The question is whether the diffusion is strong enough to 
counterbalance the spontaneous reduction of the effective dimension 
caused by the recoveries in a finite system. 
In Ref.~\cite{sakaguchi2020slow} no systematic investigation has 
been provided to understand this better. 
Furthermore, modern human societies cannot be described by regular 
Euclidean lattices, rather by small world graphs,
restricted by containment measures in the course of defense. 
Therefore the infinitely strong diffusion limit is unrealistic.
To describe epidemics usually meta-population models are used, 
built from internally strongly connected modules, which are
interconnected via diffusion of scale-free graphs~\cite{metapop}.

Here I advance another assumption for modeling societies with lock downs. 
This is to be done using hierarchical modular networks 
(HMN), where the modules can be families, villages, towns, countries 
and continents with random intra-module connections, while the modules, 
embedded in the 2d space are interconnected via long links with
geometrical distance decreasing probabilities.
In this work I investigate the dynamical behavior of SIR like models on
such HMN-s, considering hot-spot heterogeneity as well.

Another very recent publication~\cite{Mieghem-PLSIR} claims that 
empirically the average fraction of infected people decays over 
time algebraically and tries to understand it via SIR like models 
on fixed graphs.
The authors conclude that even non-Markovian description fails to explain 
the empirical data and conjecture that time-varying, human contact graphs
are needed to produce slow dynamics. Indeed, such graphs may be more 
realistic, even if containment measures freeze the mobility. 
In this work I also present results for the mobility effects on the 
fixed graph SIR model results.
 

\section{Hierarchical modular networks}

\begin{figure}[h]
\includegraphics[height=6.5cm]{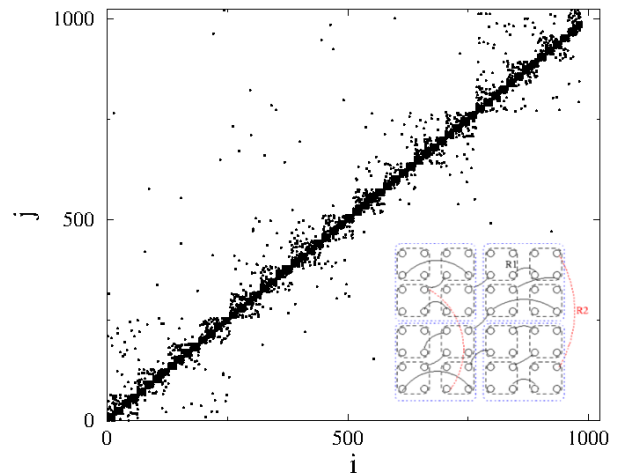}
\caption{Plot of the adjacency matrix of a $N=1024$ sized sample of the 
HMN2d graphs used for the simulations. Black dots denote edges between 
nodes $i$ and $j$. The $l_{max}=4$-level structure is clearly visible 
by the blocks near the diagonal.Low density, scattered points away from 
the diagonal represent long-range links. Inset: Scheme of the lowest 
3 levels of hierarchical, 4-block structure of nodes, embedded in the
2d space with additional long-links. Black solid lines: $l=1$ links, 
red, dashed lines $l=2$ links, $l=0$ edges are not shown.} 
\label{Amat}
\end{figure}

In this section I describe the HMN networks used for the simulations.
The network generation starts at the top level, by connecting neighbors 
to the $N$ nodes via edges with the probability 
$p_0 = b (\frac{1}{2^s})^{l_{max}}$. Here $s$ determines the type
of the network and $b$ is a control parameter. Then, further
random long links are added by level-to-level from top to bottom,
similarly as in~\cite{HMNcikk,HPTcikk}, excluding self-connections.
The levels: $l=0, 1, ...,l_{max}$ are numbered from bottom to top.
The size of domains, i.e. the number of nodes in a level, grows as
$N_l = 4^{l+1}$ in case of the $4$-module construction, related to
tiling of the 2d base lattice (see inset of Fig.~\ref{Amat}). 
The probability of random, intra-module links at level $l$ is
\begin{equation} 
p_l = b (\frac{1}{2^s})^l .
\end{equation}
With this construction he average degree
$\langle k \rangle$ of nodes is related to $b$
as shown in Table~\ref{summary}.
Nodes are connected in a hierarchical modular way as if they
were embedded in a regular, two-dimensional lattice (HMN2d) as shown by the
adjacency matrix on Fig.~\ref{Amat}, similarly as in~\cite{HMNcikk,HPTcikk}.
The $4$ nodes of the level $l=0$ are set to be fully connected.

These HMN-s, possess increasing edge density from top to bottom levels. 
Such topology has been shown to be suitable to describe activity localization 
and for the emergence of potential RR effects~\cite{KH,HMNcikk}. 
One can make a correspondence with spatially embedded networks of
type discussed in~\cite{Barthelemy}. 
These networks have long links, with algebraically decaying 
probabilities in the Euclidean distance $R$ as
\begin{equation} 
p(R) \sim R^{-s}.
\label{BB}
\end{equation}

Single connectedness of networks is not required, a typical $l_{max}=5$, 
$s=4$ sample with $N=4096$ nodes and $37240$ edges contains 3 strongly 
and 1 weakly connected components. Other, randomly selected networks showed
statistics and invariants within a few percent difference.
The modularity coefficient of the networks is high: $Q > 0.94$, defined by
\begin{equation}
Q=\frac{1}{N\av{k}}\sum\limits_{ij}\left(A_{ij}-
\frac{k_ik_j}{N\av{k}}\right)\delta(g_i,g_j),
\end{equation}
where $A_{ij}$ is the adjacency matrix and $\delta(i,j)$ is the 
Kronecker delta function.
The Watts-Strogatz clustering coefficient \cite{WS98} of a
network of $N$ nodes is
\begin{equation}\label{Cws}
C = \frac1N \sum_i 2n_i / k_i(k_i-1) \ ,
\end{equation}
where $n_i$ denotes the number of direct edges interconnecting the
$k_i$ nearest neighbors of node $i$. 
In randomly generated HMN2d-s with $N=4096$ this is roughly 
$C=0.31$, which is more than $122$ times higher than that of a 
random network of same size $C_r=0.002548$, defined by 
$C_r = \langle k\rangle / N$.
The average shortest path length is defined as
\begin{equation}
\mathcal{L} = \frac{1}{N (N-1)} \sum_{j\ne i} d(i,j) \ ,
\end{equation}
where $d(i,j)$ is the graph distance between vertices $i$ and $j$.
For several typical networks $\mathcal{L} = 10.44$ is found, which is larger 
than that of the random network of same size: 
$\mathcal{L} _r = 4$, 
computed from the formula~\cite{Fron}:
\begin{equation}
\mathcal{L} _r = \frac{\ln(N) - 0.5772}{\ln\langle k\rangle} + 1/2 \ .
\end{equation}
Hence these are small-world networks, according to the definition of
the coefficient~\cite{HumphriesGurney08}:
\begin{equation}
\sigma = \frac{C/C_r}{\mathcal{L}/\mathcal{L} _r} \ ,
\label{swcoef}
\end{equation}
because $\sigma \simeq 47$ is much larger than unity.

\begin{figure}[h]
\includegraphics[height=6.5cm]{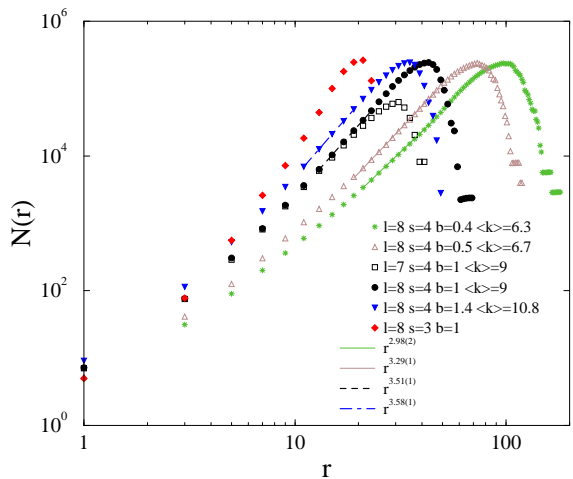}
\caption{The average number of nearest neighbors measured by the BFS 
algorithm as the function of graph distances from randomly selected
initial seeds. 
Different $b$ and $s$ values and sizes $l_{max}$ were investigated 
as shown by the legends. Lines correspond to least squared PL fits
for $10 < r < r_{cut}$, where $ r_{cut}$ was estimated visually.}
\label{dimfig}
\end{figure}

We can estimate the effective topological (graph) dimension $d_T$, 
using the breadth-first search (BFS) algorithm, defined as
\begin{equation}
N(r) \sim r^d_T \ , 
\end{equation}
by counting the number of nodes $N(r')$ with chemical distance $r'\le r$ 
within a large sample averege of trials started from randomly selected 
seeds. The dimension $d$ is estimated for different $l_{max}$, $b$ and $s$ 
values. For $s=3$ we can't find the true $d_T\to\infty$ due to the finite 
size cutoff.
As one can see on Fig.~\ref{dimfig}, for $s=4$ the graph dimension increases 
with $b$ and $\langle k\rangle$.

\section{Dynamical simulations}\label{methods}


Time dependent simulations were performed from randomly selected infected 
seed initial conditions. 
This means that at $t=0$ randomly selected $i_0$ number of nodes were set 
to the infected state: $x(i)=1$ in an otherwise fully susceptible system 
$x(j)=0$ for $j\in (1,N)$. 
Usually $i_0=1$ was used, to describe epidemics from single sources, 
but multiple source cases have also been considered.
This source triggers epidemic avalanches, used in statistical physics 
to investigate the so called critical initial slip phenomena~\cite{HHL}. 
The HMN2d graphs considered for extended simulations has $l_{max}=6,7,8$ 
levels, containing $N=16384, 65536, 262144$ nodes, respectively. 
For testing purposes $d=2$ and $d=3$ dimensional Euclidean lattices with 
linear sizes $L=100, 1000, 2000$ and periodic boundary conditions 
were also investigated.
At times $t=1,2,3, ..., t_{max}$ vector elements of the updated 
state variables are set to $x'(j)=1$, with probability $\lambda$, 
provided they were in susceptible state before $x(j)=0$ 
and had any infected neighbors.
Infected nodes recover to the state $x'(j)=-1$ with probability $\nu$.
To study mobility effects the code also performs an exchange of 
states with that of a randomly selected neighboring node, if the reaction
conditions described above are not satisfied.
Following a full sweep of nodes the old state vector is updated with
the new one : $x(j) = x'(j)$,
corresponding to one Monte Carlo step (MCs). Throughout this study I 
measure time in MCs units. Thus, Stochastic Cellular Automaton (SCA) like 
updates have been used without the loss of generality.

To prove equivalence of the scaling behavior with that of the SIR model
the simulations were tested on $d=2,3$ dimensional lattices, with periodic 
boundary conditions, for which the DIP universal scaling exponents 
are tabulated~\cite{odorbook}.
The density of infected nodes $I_m(t) = 1/N \sum_{i=1}^N \delta(x_i,1)$
is measured in each sample run $m$ and the spatio-temporal size 
$S_m = \sum_{i=1}^N \sum_{t=1}^T \delta(x_i,1)$ of the avalanches
is calculated, where $T$ denotes the maximal duration of the epidemic 
avalanche. Averaging over $I_m(t)$ of the independent samples we get
$I(t)=\langle I_m(t) \rangle$. To determine PDF of $S_m$ a histogramming 
algorithm is used on the results of thousands of realizations, started from 
random initial conditions. That means random initial infected
site locations, as well as random initial graph configurations 
in the case of the HMN2d networks. At the critical point the PL
behavior of the PDF decay tail defines the exponent 
$\tau$ as : $p(S) \propto S^{-\tau}$.
Furthermore, the avalache survival probability $P(t)$ is also determined, 
which scales at the critical point as: $P\propto t^{-\delta}$. 
To obtain more precise exponent estimates the local slopes of $I(t)$ 
is deretmined by
\begin{equation}  \label{deff}
\eta_\mathrm{eff}(t) = \frac {\ln I(t) - \ln I(t') } {\ln(t) - \ln(t')} \ ,
\end{equation}
using $t - t'= 4$. Similarly one can analyze the $P(t)$ results.
These effective exponent curves, plotted as the function of
$1/t$ veer up or veer down super- or sub-critically. At the critical point
no curvature is expected in case of simple PL sub-leading corrections to 
scaling and on can read off $\eta$ by extrapolating to $1/t \to 0$ on the
vertical axis. 

\begin{figure}[h]
\includegraphics[height=6.5cm]{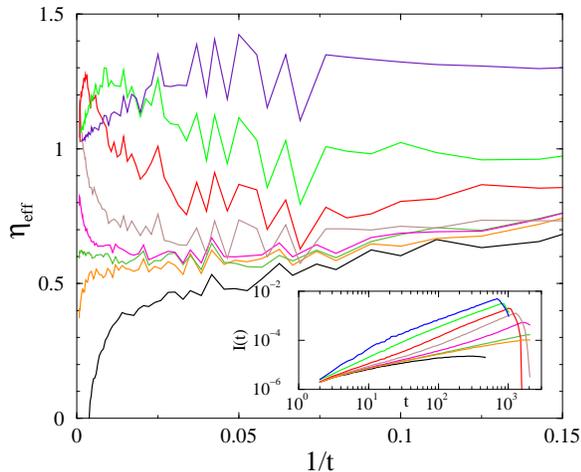}
\caption{\label{2dI}Effective exponents $\eta_\mathrm{eff}(t)$
in 2d for $\lambda=$ 0.4, 0.406, 0.407, 0.408, 0.41, 0.42, 0.44, 0.5
(bottom to top curves). 
Inset: initial time evolution of $I(t)$, averaged over runs from $10^4$ 
randomly selected initial random sites. 
The two distinct fixed point behavior can be seen at $\lambda_c=0.4059(1)$,
with $\eta=0.59(1)$ and the supercritical phase, characterized by 
$\eta = 1$.}
\end{figure}

In $d=2$ dimensions square lattices of linear size: $L=1000, 2000$ were
used and the critical scaling results were found to be in full agreement 
with those of the Dynamical Isotropic Percolation (DIP) class~\cite{HBbook,odorbook}. 
At the critical point $\lambda_c=0.4059(1)$, determined by the local
slope analysis, I estimate: $\eta=0.59(1)$ as compared to 
$\eta_{DIP2d}=0.586$~\cite{HBbook} (see Fig.\ref{2dI}). 

In $d=3$ cubes of linear sizes: $L=100, 160$ were used and at the critical
point $\lambda_c=0.2198(2)$ the growth exponent $\eta=0.53(2)$ was found 
in comparison with literature value $\eta_{DIP3d} = 0.536$~\cite{HBbook}.
Above the critical point we can observe the expected scaling,
characterized by the exponent $\eta = d-1$, before the size cutoff turns on.
However, we can also see nontrivial corrections, causing an overshoot 
of the effective exponents if $\lambda$ is slightly above $\lambda_c$. 
This correction is more pronounced in the $d=3$ case, where smaller 
sized lattices could be accessed than in $d=2$ dimensions.

\begin{figure}[h]
\includegraphics[height=6.5cm]{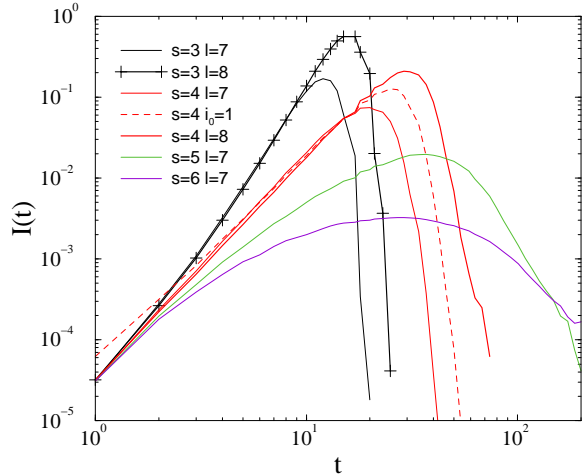}
\caption{\label{svar}Density of infected sites in different graphs 
for $i_0=2$, by varying $s$ and the size with $b=1$ and $\lambda=\nu=1$ fixed. 
Thin lines $l_{max}=7$, thick lines $l_{max}=8$ data, multiplied by a factor of 4.
Only the $s=4$ curves exhibit PL initially and exponential decay is
observable following finite size cutoff, corresponding to herd immunity.
The dashed line corresponds to the single seed case: $i_0=1$, 
multiplied by a factor 2.
}
\end{figure}

In case of the HMN2d graphs as the first step I determined the growth 
behavior of $I(t)$ at $\lambda=\nu=1$ for different $s$ values.
As one can observe on Fig.~\ref{svar}, PL-s seem to occur for $s=4$, where 
the topological dimension of the graphs is finite. 
For $s=3$ the dimension is infinite and we can see faster then PL growth 
behavior. For $s > 4$ the spatial dimension vanishes: 
$\lim_{N\to\infty }d_{T}\to 0$ and we can find slower than algebraic 
initial growth of $I(t)$. Note, that for obtaining collapse of densities 
with different sizes the $l_{max}=8$ data was multiplied by a factor of four.
I have also investigated the initial $i_0$ dependence by varying it from 
$i_0=1$ to $i_0=16$, because a recent study of the mean-field
SIR model~\cite{GinSir} suggested the possibility of $i_0$ dependent
critical exponents.
The present simulations show that the number of initial seeds scales up
the magnitude and the duration of the $I(t)$ curve, but the slope 
of the PL does not change.

\begin{figure}[h]
\includegraphics[height=6.5cm]{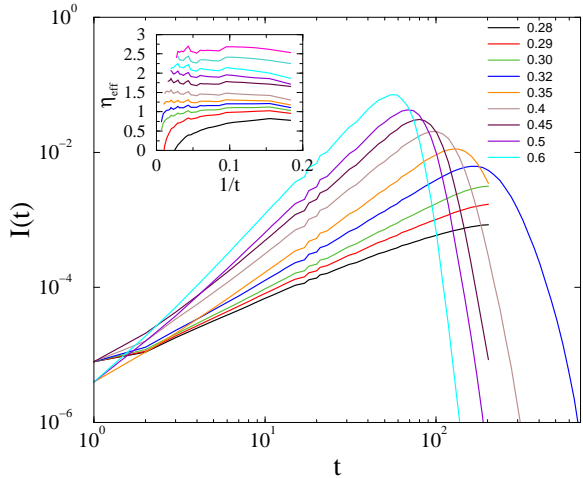}
\caption{\label{s4}Density of infected sites in graphs with $s=4$
and $b=1$ for $\lambda=$ 0.28, 0.29, 0.3, 0.32, 0.35, 0.4, 0.45, 0.5, 0.6 
(bottom to top curves). Inset: local slopes of the same curves as well as
for $\lambda=$0.7, 0.8. One may think of continuously changing exponents 
above $\lambda_c=0.310(5)$, because short times hinder to see the 
supercritical point scaling behavior with $\eta\simeq 2.5$.}
\end{figure}

Next I concentrated on the $s=4$, $b=1$ case and determined the infection
probability dependence as shown on Fig.~\ref{s4} using $\nu=1-\lambda$. 
In this case one initially thinks of $\lambda$ dependent initial PL-s
by looking at the results.
When we calculate the local slopes of $I(t)$ we can observe
a phase transition point at $\lambda_c \simeq 0.3$ characterized by 
$\eta=1.4(1)$ as it appears in the inset of Fig.~\ref{s4}. 
This exponent is much bigger than that of the DIP universality
class value for the $d=3$ case: $\eta_{DIP3d} = 0.536(10)$~\cite{Munoz99}. 
Since $\eta_{DIP}$ decreases further, when the spatial dimension 
increases the HMN2d exponent is in conflict with the expectation of a 
homogeneous system projection for $d_T=3.5$, obtained by BFS analysis
for this network.
Thus, the topological disorder alters the critical point 
scaling behavior.

We can see that the supercritical curves with $\lambda_c < \lambda < 0.45$ 
veer up, while those with $\lambda > 0.45$ veer down, converging to 
$\eta_{s} \simeq 2.5(10)$, in agreement with the result obtained by
the BFS dimension measurements, according to which
we expect $\eta_s = d_T - 1 \simeq 2.5$. Note, that the small variation
of the effective exponents in the narrow scaling region before the
finite size cutoff may suggest the wrong conclusion of $\lambda$ 
dependent scaling exponents.
We have no reason to believe in such non-universal exponents here in the
lack of long surviving RR-s. This will be more obvious later,
at the $b=0.4$ case, where the exponents of the two fixed point, the 
critical and the supercritical one are more distant.
I have also performed test runs by starting the system from homogeneous,
$80\%$ randomly infected initial states near criticality 
and found simple exponential decays of $I(t)$, ruling out of a 
GP like behavior. Note that the small zig-zags at $t\simeq 20$ are
numerical artifacts, coming from the combination of SCA updating on 
HMN2d lattices. They do not appear in the regular lattice simulations.

\begin{figure}[h]
\includegraphics[height=6.5cm]{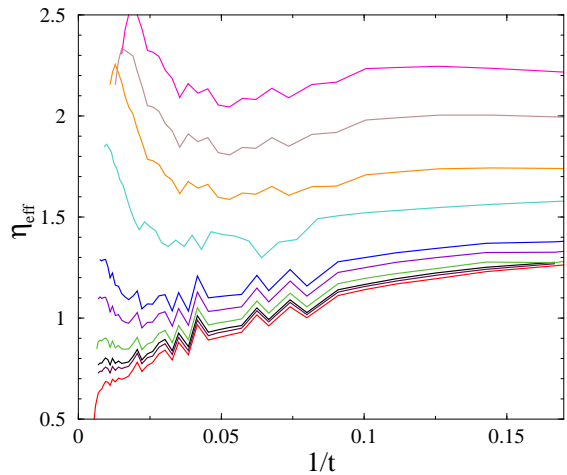}
\caption{\label{b04} Effective exponents $\eta_\mathrm{eff}$
as in Fig.~\ref{s4}, for $s=4$ and $b=0.4$ for $\lambda$= 
0.47, 0.473, 0.475, 0.48, 0.49, 0.5, 0.55, 0.6, 0.7, 0.8 (bottom to top curves).
The two distinct fixed point behavior can be seen at $\lambda_c=0.480(5)$,
with $\eta=0.8(1)$ and the supercritical phase, characterized by
$\eta \simeq 2$.}
\end{figure}

I have repeated these simulations for other $b$ values at $s=4$
and found similar results. For example at $b=0.4$ the average
degree is $\langle k\rangle = 6.3$ and the topological dimension is
$d_T \simeq 3$. The curves of the effective exponents on Fig.~\ref{b04}
veer down for $\lambda < 0.48$ and veer up for $\lambda > 0.48$
as $t \to \infty$, but finite size terminates the epidemic and causes
an exponential cutoff.
At $\lambda_c \simeq 0.475(1)$ we can read off an asymptotic PL
growth behavior characterized by the exponent $\eta=0.8(1)$, much 
larger than the homogeneous, 3 dimensional system value: 
$\eta_{DIP3d}=0.53(2)$.
One can also see clearly, that in the supercritical phase the effective
exponent curves do not level off, but tend to $\eta \simeq 2 = d_T-1$,
following the overshoot correction region.
For other $b$ values we can obtain similar results, the critical and the
supercritical $\eta$ exponents change continuously as shown in 
Table~\ref{summary}, summarizing the results.

\begin{table}
 \caption{\label{summary} Summary of critical SIR results for $s=4$ HMN2d
  networks and Euclidean lattices. 
 The type of graph is described by the Euclidean dimension (2d, 3d),
 or by the value of $b$ for HMN2d. "$+D$" denotes the diffusive case, 
 "$+H$" means the application of a single super-spreader hot-spot.}
 \centering
 \begin{tabular}{|c|c|c|c|c|c|}
  \hline
   Type  & $\langle k \rangle$  & $\lambda_c$  & $\eta$ & $\tau$   & $d_T$ \\
   \hline
      2d &      4              & 0.4059(1)    & 0.59(1) & 1.06(1) & 2 \\
      3d &      6	       & 0.2198(2)    & 0.53(2) & 1.20(2) & 3 \\  			
   \hline
   2d+D  &      4              & 0.4135(1)    & 0.25(2) & 1.05(1) & 2 \\ 
    \hline
    \hline
    $b=0.4$ &  6.3             & 0.475(1)     & 0.8(1) & 1.05(5) & 2.98(2) \\
    $b=0.5$ &  6.7	       & 0.425(5)     & 0.95(4) & 1.01(3) & 3.29(1)\\
    $b=1.0$ &  9.1	       & 0.310(5)     & 1.4(1) & 1.10(7) & 3.5(1)\\	
    $b=1.5$ &  9.3	       & 0.23(1)      & 1.30(3)  & 1.12(5) & 3.8(1)\\ 
\hline
   $b=1.0+D$&      9.1             & 0.240(3)     & 1.4(1)  & 1.10(8) & 3.5(1)\\
 \hline
  $b=1.0+D+H$&     9.1             & 0.241(3)     & 1.4(1)  & 1.11(5) & 3.5(1)\\
 \hline

 \end{tabular}
\end{table}

\subsection{Size distribution of epidemics}

The total epidemic size statistics, determined for different countries, 
also show PL distributions~\cite{doi:10.1063/5.0013031} 
and a snowball model on a two-level, heterogeneous system was suggested 
to describe it. 
Another, earlier, brain motivated study of a SIR like system, 
applied on HMN2d graphs also concluded PL-s for the
fractional component sizes~\cite{doi:10.1063/1.4793782} using
normalization and simulation methods.

I have determined the PDF-s of the infection spatio-temporal avalanche 
sizes of the $s=4$ case for different $b$ and $\lambda$ values near
criticality. These distributions seem to exhibit PL tails before a bump 
at the end in case of supercriticality, corresponding to a giant component.
Again, first I calculated these distributions for testing in case of 
Euclidean lattices and found results in good agreement with the corresponding 
critical DIP classes. In particular, in $d = 2$ I obtained: $\tau= 1.06(1)$
as compared to $\tau_{DIP2d}=96/91 \simeq 1.055$ of the DIP 
class~\cite{HBbook}. In $d=3$ this model provides $\tau= 1.20(2)$ with
respect to $\tau_{DIP3d} = 1.188$ of the DIP class~\cite{HBbook}.
For the supercritical $\lambda$-s in infinite systems the epidemic 
never stops, thus the  $p(s)$ distribution is singular:  
$p(S) \propto S^{-1}$.

\begin{figure}[h]
\includegraphics[height=6.5cm]{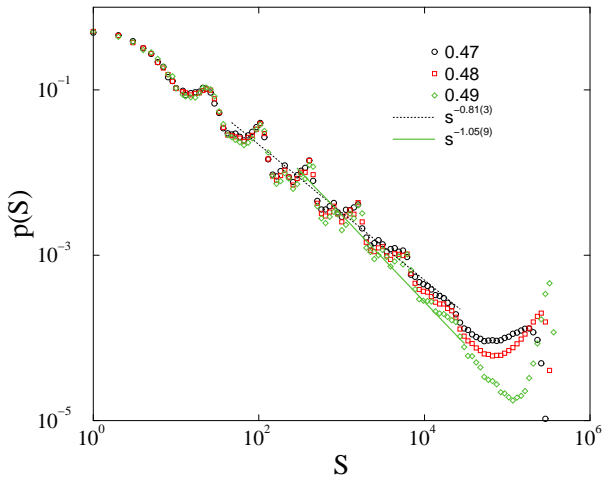}
\caption{\label{avs4} Spatio-temporal size distribution of 
infection avalanches for HMN2d-s with $s=4$ and $b=0.4$ at
different $\lambda$-s and $\nu=1-\lambda$. Lines present PL fits for
the $s>10$ region before the bump.}
\end{figure}

Fig.~\ref{avs4} shows the results for HMN2d-s with $s=4$ and $b=0.4$, 
corresponding to average degree $\langle k \rangle= 6.3$ and graph 
dimension $d_T \simeq 3$. A PL fit for the intermediate "tail" region, 
which is $s > 10$ before the bump, corresponding to a giant component,
results in: $\tau=1.05(5)$, a slightly smaller exponent than that of 
the regular lattice: $\tau_{DIP3d} = 1.20(2)$. Several bumps
can be seen, due to modules of different sizes. These log-periodic 
oscillations, superimposed on the PL-s are the consequence of the
discrete scale invariance of the HMN2d graph and increase the numerical 
uncertainty of the fitting procedure.

\subsection{The effect of mobility}

To simulate time-varying, human contact graphs I repeated the 
aforementioned spreading analysis for $2d$ lattices as well as for 
the HMN2d-s with $s=4$, by allowing the diffusion of states. 
This means that SIR individuals are not fixed and can have different
neighbors. The simulation program emulate this by an additional
state exchange of $x(i)$ with the state of a randomly selected 
neighbor $x(j)$, provided the reaction conditions are not satisfied. 

\begin{figure}[h]
\includegraphics[height=6.5cm]{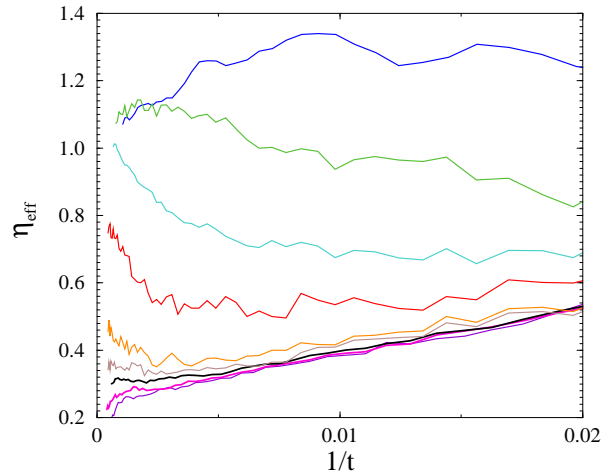}
\caption{\label{2dD} The effect of diffusion on the local slopes
of $I(t)$ in 2d lattices of sizes $L=1000,2000$ for different 
$\lambda=$ 0.4125, 0.413, 0.4135, 0.414, 0.415, 0.42, 0.43, 0.45, 0.5
(for curves from bottom to top). $L=2000$ results are plotted near
the critical point: $\lambda=0.413, 0.4135$ (thinner lines). }
\end{figure}

For the $2d$ lattice the supercritical scaling is invariant, but
the critical point increases slightly to $\lambda_c=0.413(1)$ and
the growth exponents decreases to $\eta=0.25(2)$ (see Fig.\ref{2dD}),
deviating considerably from the 2d SIR exponent.
This is the consequence of the site reinfecibility, the
long term memory of sites is lost, causing SIS type of critical
behavior. This exponent is close to the 2d DP universality 
class value: $\eta_{2d,DP}=0.2295(10)$~\cite{marro2005,HHL,odorbook}.
It is hard to determine it very precisely, due to the finite size
cutoff, but increasing the size from $L=1000$ to $L=2000$ just
above $\lambda_c$ one can observe a up-bend curvature before the
cutoff, which  moves the estimate towards $\eta_{2d,DP}$.
However, for the survival probabilty exponent we can obtain an
estimate: $\delta=0.20(1)$ before the exponential break down, 
which is far away from that of the DP value $\delta_{2d,DP}=0.4505(10)$ 
and also from the DIP value $\delta_{2d,DIP}=0.092$
\cite{marro2005,HHL,odorbook}.
Interestingly the diffusion increases the inactive phase, 
susceptible individuals can diffuse back, behind the growing epidemic 
front, becoming target of re-infection.
The size distribution exponent $\tau$ does not change, neither
the exponential decay in the herd immunity phase.

\begin{figure}[h]
\includegraphics[height=6.5cm]{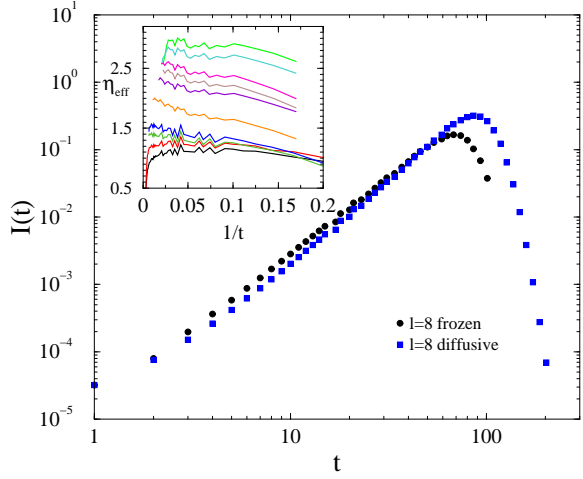}
\caption{\label{ss} The effect of diffusion on $I(t)$ in HMN2d graphs 
with $s=4$, $b=1$ and $\lambda=\nu=0.5$ fixed. 
The diffusion (blue boxes) increases 
the size of epidemic, but does not alter the exponents.
Inset: Local slopes of $I(t)$ in HMN2d graphs with $s=4$, $b=1$, for 
$\lambda=$ 0.22, 0.23, 0.24, 0.25, 0.30, 0.37, 0.44, 0.5, 0.6, 0.7 
(bottom to top curves) in the presence of diffusion. 
The critical and supercritical $\eta$ are the same as without diffusion.}
\end{figure}

In case of the HMN2d networks the diffusion does not seems to
change the dynamical exponents. 
As we can see on Fig.\ref{ss}, in the supercritical phase 
at $\lambda=0.5$, the initial scaling behavior remains the same, 
characterized by $\eta \simeq d_T - 1 = 2.5$, but the epidemic grows 
further, achieving
a larger maximum value than in the frozen case. The inset of
The inset of Fig.~\ref{ss} shows the effective growth exponent
results for different $\lambda$-s at $b=1$ near the critical 
point. Again a critical point at $\lambda_c = 0.240(3)$ appears,
lower than that of the frozen case (0.310(5)) and in supercritical
phase $\eta_{eff}$ estimates tend to  $\eta\simeq 2.5$.

\begin{figure}[h]
\includegraphics[height=6.5cm]{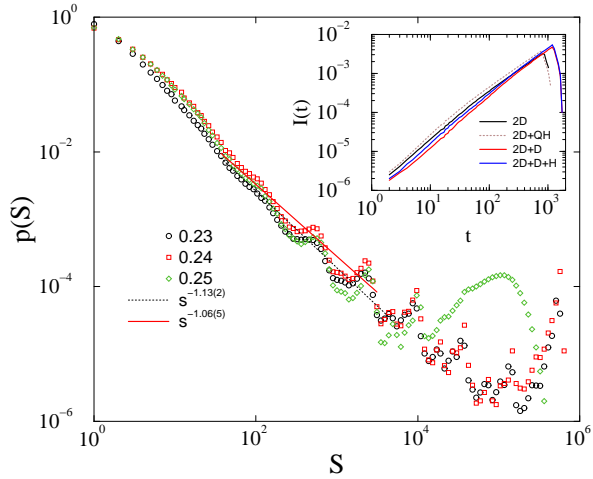}
\caption{\label{avs4D} Spatio-temporal size distribution of infection
avalanches for HMN2d-s at $s=4$ and $b=1$ in the presence of diffusion
for different $\lambda$= 0.23 (circles), 0.24 (squares), 0.25 (rhombes). 
Lines present PL fits for the $s>10$ region before the bump. 
Inset: The effect of hot-spots (H) in 2d at $\lambda=0.5$. 
Brown, dashed line: $10 \%$ quenched, random distributed H-s,
black, solid line: homogenous 2d, blue line: diffusion (D) + H,
red line D only (top to bottom curves).
}
\end{figure}

The scaling behavior of the epidemic size distributions $P(S)$ exhibit 
insenitivity to the mobility as shown by main plot of Fig.~\ref{avs4D}),
but the oscillations are even more pronounced.
The fitted exponent for the tails before the bump, corresponding to the 
giant component $\tau=1.10(8)$ is in good agreement with that of
the frozen network case.

In contrast with the 2d lattice, the mobility does not seem to alter
the critical point scaling, probably because the strongly 
connected network structure does not allow recovery, keeping the 
long-time local memory of sites.

\subsection{Super-spreader hot spot in the presence of mobility}

\begin{figure}[h]
\includegraphics[height=6.5cm]{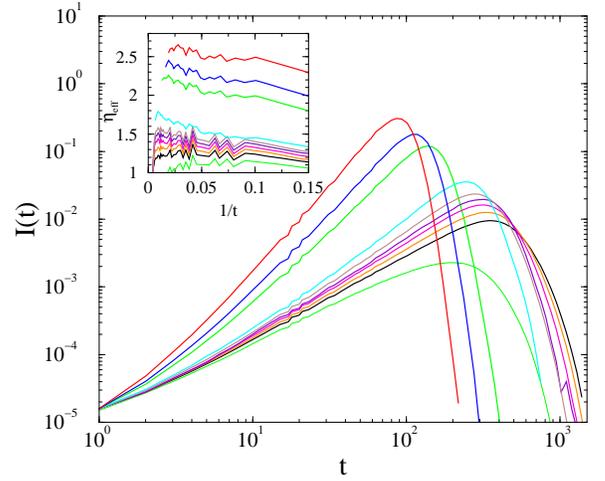}
\caption{\label{w1} The effect of a single hot-spot for the diffusive model
in HMN2d graphs with $s=4$, $b=1$ and $\lambda=0.22$, 0.23, 0.235, 0.24, 0.245,
0.25, 0.26, 0.4, 0.5 (bottom to top curves). 
Inset: Local slopes of the same. The asymptotic critical and supercritical
effective exponents are roughly the same as in the non-diffusive homogeneous
SIR.}
\end{figure}

I have also investigated the effect of a super-spreader hot spot in the 
presence of mobility as suggested in Ref.~\cite{sakaguchi2020slow}. 
This was achieved by increasing the infection probability at a single 
site $i=100$ to $\lambda = 1$.
However, this change alone did not cause measurable difference in the 
epidemic, neither in the initial regime, nor in the size distributions
in case of HMN2d-s. 
Fig.~\ref{w1} shows the $I(t)$ for $s=4$ and $b=1$ situation.
In the herd immunity regime the decay remains exponential.
Thus in the HMN2d graphs, where the epidemic propagation is restricted 
by the "containment" of modules neither the mobility nor hot-spots
can lead to slow decay dynamics as in Ref.~\cite{sakaguchi2020slow}.

In case of 2d diffusive lattices at $\lambda=0.5$ this single hot-spot 
increased the size of $I(t)$, by growing the maximum by $20 \%$, as
one can observe in the inset of Fig.~\ref{avs4D}, but the decay 
remains exponential. Without diffusion, quenched hot-spots do
not have strong effects in 2d, even a $10 \%$ concentration
of randomly distributed ones cause similar growth with respect
to the homogenous lattice and the peak time seems to decrease.
The details of quenched disorder runs are presented in the Appendix.


\section{Conclusions}


SCA models following SIR rules on regular lattices and on
hierarchical modular graphs have been investigated by numerical
simulations. Scaling behavior at the epidemic outbreak 
was found in systems with finite topological dimensions. 
The dynamical percolation behavior on $d=2,3$ dimensional lattices 
was fully confirmed and a nontrivial correction to scaling 
in the supercritical phase has been explored in detail.

To model human society with containment measures hierarchical modular
graphs, embedded in 2d space (HMN2d) were used. 
A special set of graphs were considered, in which long-range connections 
decay with the geometrical distance in a PL manner and the 
topological dimension, together with the average degree can be 
tuned via a single parameter.
I found that the critical behavior is altered by the topological
heterogeneity, such that $\langle k\rangle$ dependent exponents 
occur for the number of newly infected agents as well as for the
total spatio-temporal sizes of pandemics, which can be regarded
avalanches, triggered by a single infected site. By changing
continuously $\langle k\rangle$, non-universal PL behavior 
emerges for the growth and epidemic sizes, characterized by 
the continuously varying exponents $\eta$ and $\tau$.
Comparing exponents of a $d_T\simeq 3$ HMN2d with those of a $d=3$ 
Euclidean lattice provides different exponents, thus it turns out that
the topological heterogeneity is relevant for modifying the 
dynamical scaling behavior. 

On the other hand, the supercritical scaling is insensible to
the heterogeneity, one can find the same growth laws as in
the corresponding $d$ dimensional Euclidean lattices.
The epidemic size distributions also exhibit PL tail region 
at criticality, with exponents smaller, but close to those of 
the homogeneous lattices.
They increase slightly with $\langle k\rangle$.

Note, that as the scaling region, before herd immunity sets in, 
is narrow in finite systems and one can misleadingly think of $\lambda$ 
dependent exponents for a given topology. Empirical data may also 
suggest this. Only a detailed, local slope scaling analysis,
showing the scaling corrections allows one to determine the 
true asymptotic behavior. 
Smaller system sizes make the outbreak scaling region narrower 
as well as the maximum of $I(t)$ smaller, but do not change the 
scaling exponents. Multiple sources also do not affect 
the exponents, but the scale up sizes of the epidemics. 

Comparing values of $\eta$ and $\tau$ with those from
COVID-19 statistics we can see that a proper HMN2d model 
description would require larger $\langle k\rangle$ and 
graph dimensions to obtain agreement with real data.
Fitting for the confirmed COVID-19 case statistics
$\tau = 1.14 - 1.5$ exponents are reported in~\cite{doi:10.1063/5.0013031}.
The largest critical value investigated here was
$\tau = 1.12(5)$ but this can grow further by increasing the 
control parameter $b$. 
Of course the epidemics are not necessarily critical, nor their 
connection network is finite dimensional.
Furthermore, in reality the parameters change in time, thus
for example the catastrophic case with a giant component 
can be avoided. 

Effects of mobility and the possibility of hot-spots with 
large local infection rates have also been investigated.
Diffusion washes out long term memory of sites and in the 2d 
lattice and we can observe DP universality class like initial 
slip scaling exponent $\eta$ as in case of the memoryless SIS model.
However, the avalanche survival exponent $\delta$ is different.
Further investigation to clarify if this is a new universality
class or diffusion strength dependence occurs would be needed. 

In case of HMN2d the mobility did not change the late time 
exponential decay dynamics. It increases the size of epidemics, 
but does not seem to alter the critical exponents. 
Therefore, the HMN2d modular topology is efficient to keep the 
functional form of the frozen SIR epidemic dynamics.
The addition of a single hot-spot also turned out to be irrelevant
both for the initial and for long time epidemic behaviors.

In case of the 2d lattice model the diffusion had the effect of 
increasing the maximum of $I(t)$ by $\simeq 20\%$ and doubling the 
duration with respect to the pure system at $\lambda=0.5$.
Further more detailed studies would be needed to clarify the hot-spot 
effects, involving a comparison with random sequential update 
model simulations.

It would be also be an interesting extension of research to clarify
the effects of the intrinsic quenched disorder on the SIR dynamics 
as Harris criterion predicts irrelevance. Preliminary runs in
$d = 2$ Euclidean lattices with $10\%$ of randomly added quenched 
hot-spots do not modify the critical dynamics (see Appendix). 
However, the combination of hot-spots with mobility may affect
the initial growth scaling more profoundly.

I thank R\'obert Juh\'asz for the discussions and Silvio
Ferreira for the useful comments.
Support from the Hungarian National Research, Development 
and Innovation Office NKFIH (K128989) is acknowledged. 
I thank access to the Hungarian National Supercomputer Network.

\bigskip

\section*{Appendix}

\begin{figure}[h]
\includegraphics[height=6.5cm]{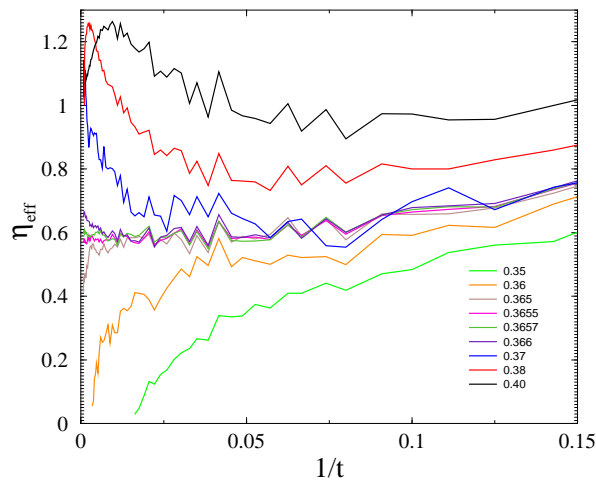}
\caption{\label{A1} Effective exponents $\eta_\mathrm{eff}$
as in Fig.~\ref{s4} for SIR on the $d=2$ lattices, with bi-modal
quenched disorder for $\lambda$= 0.35, 0.36, 0.365, 0.3655, 0.3557, 
0.366, 0.37, 0.38, 0.40 (bottom to top curves).}
\end{figure}

\begin{figure}[h]
\includegraphics[height=6.5cm]{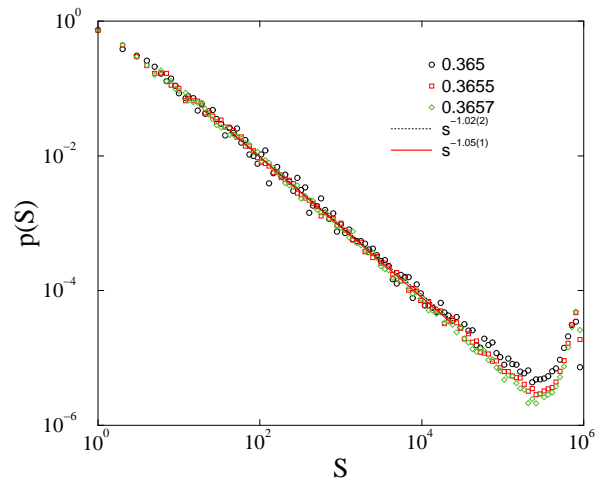}
\caption{\label{avs2Dq} Spatio-temporal size distribution of infection
avalanches in 2d in the presence of bi-modal quenched disorder.
for different $\lambda$-s and $\nu=1-\lambda$. Lines present PL fits for
the $s>10$ region before the bump.}
\end{figure}

In this Appendix I show some preliminary results for the study of
interaction disorder in case of homogeneous, $d=2$ Euclidean lattices
of size $L=1000$. The simulation methods are the same as discussed in
Section~\ref{methods}. I investigated heterogeneity in the form of
quenched, intrinsic disorder. This was done by initializing the system 
with heterogeneous $\lambda_i$-s. I used bi-modal disorder distribution 
\begin{equation}\label{bimodal}
P(\lambda_i) = (1-q) \delta(\lambda_i - \lambda) + q \delta(\lambda_i - 1) \ \ ,
\end{equation}
with $q=0.1$. This means, that at randomly selected $10\%$ of sites
maximum infection probability ($\lambda_i=1$), super-spreader spots are 
created. 
Again, averaging over thousands of samples we can determine the critical 
point, which is lower than that of the pure case: $\lambda_c=0.3656(1)$, 
and one can read off the same $\eta=0.59(1)$ growth exponent
as in case of the homogeneous system (see Fig.~\ref{A1}). The $10\%$ decrease
of $\lambda_c$, compared to the homogeneous system means a substantial
growth of the endemic phase as well as the epidemic sizes
due to the super-spreader sites. 
The scaling in the supercritical behavior is unaltered, it follows 
the linear growth asymptotically.
The scaling of size distributions also agree with that of the pure
case, with $\tau=1.05(1)$ as can we see on Fig.~\ref{avs2Dq}.
Therefore, at least "weak" quenched interaction disorder seems to be
irrelevant for the SIR critical behavior, not breaking the Harris 
criterion prediction.

\bibliography{hybGP}

\begin{thebibliography}{23}
\expandafter\ifx\csname natexlab\endcsname\relax\def\natexlab#1{#1}\fi
\expandafter\ifx\csname bibnamefont\endcsname\relax
  \def\bibnamefont#1{#1}\fi
\expandafter\ifx\csname bibfnamefont\endcsname\relax
  \def\bibfnamefont#1{#1}\fi
\expandafter\ifx\csname citenamefont\endcsname\relax
  \def\citenamefont#1{#1}\fi
\expandafter\ifx\csname url\endcsname\relax
  \def\url#1{\texttt{#1}}\fi
\expandafter\ifx\csname urlprefix\endcsname\relax\def\urlprefix{URL }\fi
\providecommand{\bibinfo}[2]{#2}
\providecommand{\eprint}[2][]{\url{#2}}

\bibitem[{\citenamefont{Pastor-Satorras
  et~al.}(2015)\citenamefont{Pastor-Satorras, Castellano, {Van Mieghem}, and
  Vespignani}}]{PastorSatorras2015}
\bibinfo{author}{\bibfnamefont{R.}~\bibnamefont{Pastor-Satorras}},
  \bibinfo{author}{\bibfnamefont{C.}~\bibnamefont{Castellano}},
  \bibinfo{author}{\bibfnamefont{P.}~\bibnamefont{{Van Mieghem}}},
  \bibnamefont{and}
  \bibinfo{author}{\bibfnamefont{A.}~\bibnamefont{Vespignani}},
  \bibinfo{journal}{Reviews of Modern Physics} \textbf{\bibinfo{volume}{87}},
  \bibinfo{pages}{925} (\bibinfo{year}{2015}).

\bibitem[{\citenamefont{Marro and Dickman}(2005)}]{marro2005}
\bibinfo{author}{\bibfnamefont{J.}~\bibnamefont{Marro}} \bibnamefont{and}
  \bibinfo{author}{\bibfnamefont{R.}~\bibnamefont{Dickman}},
  \emph{\bibinfo{title}{Nonequilibrium Phase Transitions in Lattice Models}},
  Al{\'e}a-Saclay (\bibinfo{publisher}{Cambridge University Press},
  \bibinfo{year}{2005}), ISBN \bibinfo{isbn}{9780521019460}.

\bibitem[{\citenamefont{Henkel et~al.}(2008)\citenamefont{Henkel, Hinrichsen,
  and L\"ubeck}}]{HHL}
\bibinfo{author}{\bibfnamefont{M.}~\bibnamefont{Henkel}},
  \bibinfo{author}{\bibfnamefont{H.}~\bibnamefont{Hinrichsen}},
  \bibnamefont{and} \bibinfo{author}{\bibfnamefont{S.}~\bibnamefont{L\"ubeck}},
  \emph{\bibinfo{title}{Non-equilibrium phase transition: Absorbing Phase
  Transitions}} (\bibinfo{publisher}{Springer Verlag},
  \bibinfo{address}{Netherlands}, \bibinfo{year}{2008}).

\bibitem[{\citenamefont{{\'O}dor}(2008)}]{odorbook}
\bibinfo{author}{\bibfnamefont{G.}~\bibnamefont{{\'O}dor}},
  \emph{\bibinfo{title}{Universality in nonequilibrium lattice systems:
  Theoretical foundations}} (\bibinfo{publisher}{World Scientific},
  \bibinfo{year}{2008}).

\bibitem[{\citenamefont{Mu{\~n}oz et~al.}(1999)\citenamefont{Mu{\~n}oz,
  Dickman, Vespignani, and Zapperi}}]{Munoz99}
\bibinfo{author}{\bibfnamefont{M.~A.} \bibnamefont{Mu{\~n}oz}},
  \bibinfo{author}{\bibfnamefont{R.}~\bibnamefont{Dickman}},
  \bibinfo{author}{\bibfnamefont{A.}~\bibnamefont{Vespignani}},
  \bibnamefont{and} \bibinfo{author}{\bibfnamefont{S.}~\bibnamefont{Zapperi}},
  \bibinfo{journal}{Phys. Rev. E} \textbf{\bibinfo{volume}{59}},
  \bibinfo{pages}{6175} (\bibinfo{year}{1999}).

\bibitem[{\citenamefont{Harris}(1974)}]{Harris}
\bibinfo{author}{\bibfnamefont{A.~B.} \bibnamefont{Harris}},
  \bibinfo{journal}{Journal of Physics C: Solid State Physics}
  \textbf{\bibinfo{volume}{7}}, \bibinfo{pages}{1671} (\bibinfo{year}{1974}).

\bibitem[{\citenamefont{Grassberger}(1983)}]{gras83}
\bibinfo{author}{\bibfnamefont{P.}~\bibnamefont{Grassberger}},
  \bibinfo{journal}{Mathematical Biosciences} \textbf{\bibinfo{volume}{63}},
  \bibinfo{pages}{157} (\bibinfo{year}{1983}), ISSN \bibinfo{issn}{0025-5564}.

\bibitem[{\citenamefont{Vojta}(2006)}]{Vojta2006b}
\bibinfo{author}{\bibfnamefont{T.}~\bibnamefont{Vojta}},
  \bibinfo{journal}{Journal of Physics A: Mathematical and General}
  \textbf{\bibinfo{volume}{39}}, \bibinfo{pages}{R143} (\bibinfo{year}{2006}).

\bibitem[{\citenamefont{Komarova et~al.}(2020)\citenamefont{Komarova, Schang,
  and Wodarz}}]{doi:10.1098/rsif.2020.0518}
\bibinfo{author}{\bibfnamefont{N.~L.} \bibnamefont{Komarova}},
  \bibinfo{author}{\bibfnamefont{L.~M.} \bibnamefont{Schang}},
  \bibnamefont{and} \bibinfo{author}{\bibfnamefont{D.}~\bibnamefont{Wodarz}},
  \bibinfo{journal}{Journal of The Royal Society Interface}
  \textbf{\bibinfo{volume}{17}}, \bibinfo{pages}{20200518}
  (\bibinfo{year}{2020}).

\bibitem[{\citenamefont{Sakaguchi and Nakao}(2021)}]{sakaguchi2020slow}
\bibinfo{author}{\bibfnamefont{H.}~\bibnamefont{Sakaguchi}} \bibnamefont{and}
  \bibinfo{author}{\bibfnamefont{Y.}~\bibnamefont{Nakao}},
  \bibinfo{journal}{Phys. Rev. E} \textbf{\bibinfo{volume}{103}},
  \bibinfo{pages}{012301} (\bibinfo{year}{2021}).

\bibitem[{\citenamefont{Colizza and A.}(2008)}]{metapop}
\bibinfo{author}{\bibfnamefont{V.}~\bibnamefont{Colizza}} \bibnamefont{and}
  \bibinfo{author}{\bibfnamefont{V.}~\bibnamefont{A.}},
  \bibinfo{journal}{Journal of Theoretical Biology}
  \textbf{\bibinfo{volume}{251}}, \bibinfo{pages}{450} (\bibinfo{year}{2008}),
  ISSN \bibinfo{issn}{0022-5193}.

\bibitem[{\citenamefont{Mieghem P.~Van and Q.}(2020)}]{Mieghem-PLSIR}
\bibinfo{author}{\bibfnamefont{A.~M.} \bibnamefont{Mieghem P.~Van}}
  \bibnamefont{and} \bibinfo{author}{\bibfnamefont{L.}~\bibnamefont{Q.}},
  \emph{\bibinfo{title}{Power-law decay in epidemics is likely due to
  interactions with the time-variant contact grap}} (\bibinfo{year}{2020}),
  \eprint{Delft University of Technology, report20201201}.

\bibitem[{\citenamefont{{\'O}dor et~al.}(2015)\citenamefont{{\'O}dor, Dickman,
  and {\'O}dor}}]{HMNcikk}
\bibinfo{author}{\bibfnamefont{G.}~\bibnamefont{{\'O}dor}},
  \bibinfo{author}{\bibfnamefont{R.}~\bibnamefont{Dickman}}, \bibnamefont{and}
  \bibinfo{author}{\bibfnamefont{G.}~\bibnamefont{{\'O}dor}},
  \bibinfo{journal}{Scientific Reports} \textbf{\bibinfo{volume}{5}},
  \bibinfo{pages}{14451} (\bibinfo{year}{2015}).

\bibitem[{\citenamefont{\'Odor and de~Simoni}(2021)}]{HPTcikk}
\bibinfo{author}{\bibfnamefont{G.}~\bibnamefont{\'Odor}} \bibnamefont{and}
  \bibinfo{author}{\bibfnamefont{B.}~\bibnamefont{de~Simoni}},
  \bibinfo{journal}{Phys. Rev. Research} \textbf{\bibinfo{volume}{3}},
  \bibinfo{pages}{013106} (\bibinfo{year}{2021}).

\bibitem[{\citenamefont{Kaiser and Hilgetag}(2010)}]{KH}
\bibinfo{author}{\bibfnamefont{M.}~\bibnamefont{Kaiser}} \bibnamefont{and}
  \bibinfo{author}{\bibfnamefont{C.}~\bibnamefont{Hilgetag}},
  \bibinfo{journal}{Frontiers in Neuroinformatics} \textbf{\bibinfo{volume}{4}}
  (\bibinfo{year}{2010}).

\bibitem[{\citenamefont{Barthelemy}(2018)}]{Barthelemy}
\bibinfo{author}{\bibfnamefont{M.}~\bibnamefont{Barthelemy}},
  \bibinfo{journal}{Comptes Rendus Physique} \textbf{\bibinfo{volume}{19}},
  \bibinfo{pages}{205–232} (\bibinfo{year}{2018}), ISSN
  \bibinfo{issn}{1631-0705}.

\bibitem[{\citenamefont{Watts and Strogatz}(1998)}]{WS98}
\bibinfo{author}{\bibfnamefont{D.~J.} \bibnamefont{Watts}} \bibnamefont{and}
  \bibinfo{author}{\bibfnamefont{S.~H.} \bibnamefont{Strogatz}},
  \bibinfo{journal}{Nature} \textbf{\bibinfo{volume}{393}},
  \bibinfo{pages}{440} (\bibinfo{year}{1998}), ISSN \bibinfo{issn}{1476-4687}.

\bibitem[{\citenamefont{Fronczak et~al.}(2004)\citenamefont{Fronczak, Fronczak,
  and Ho\l{}yst}}]{Fron}
\bibinfo{author}{\bibfnamefont{A.}~\bibnamefont{Fronczak}},
  \bibinfo{author}{\bibfnamefont{P.}~\bibnamefont{Fronczak}}, \bibnamefont{and}
  \bibinfo{author}{\bibfnamefont{J.~A.} \bibnamefont{Ho\l{}yst}},
  \bibinfo{journal}{Phys. Rev. E} \textbf{\bibinfo{volume}{70}},
  \bibinfo{pages}{056110} (\bibinfo{year}{2004}).

\bibitem[{\citenamefont{Humphries and Gurney}(2008)}]{HumphriesGurney08}
\bibinfo{author}{\bibfnamefont{M.~D.} \bibnamefont{Humphries}}
  \bibnamefont{and} \bibinfo{author}{\bibfnamefont{K.}~\bibnamefont{Gurney}},
  \bibinfo{journal}{PLOS ONE} \textbf{\bibinfo{volume}{3}}, \bibinfo{pages}{1}
  (\bibinfo{year}{2008}).

\bibitem[{\citenamefont{Bunde and Havlin}(1991)}]{HBbook}
\bibinfo{author}{\bibfnamefont{A.}~\bibnamefont{Bunde}} \bibnamefont{and}
  \bibinfo{author}{\bibfnamefont{S.}~\bibnamefont{Havlin}},
  \emph{\bibinfo{title}{Fractals and Disordered Systems}}
  (\bibinfo{publisher}{Springer Verlag}, \bibinfo{address}{Heidelberg},
  \bibinfo{year}{1991}).

\bibitem[{\citenamefont{Radicchi and Bianconi}(2020)}]{GinSir}
\bibinfo{author}{\bibfnamefont{F.}~\bibnamefont{Radicchi}} \bibnamefont{and}
  \bibinfo{author}{\bibfnamefont{G.}~\bibnamefont{Bianconi}},
  \bibinfo{journal}{Phys. Rev. E} \textbf{\bibinfo{volume}{102}},
  \bibinfo{pages}{052309} (\bibinfo{year}{2020}).

\bibitem[{\citenamefont{Blasius}(2020)}]{doi:10.1063/5.0013031}
\bibinfo{author}{\bibfnamefont{B.}~\bibnamefont{Blasius}},
  \bibinfo{journal}{Chaos: An Interdisciplinary Journal of Nonlinear Science}
  \textbf{\bibinfo{volume}{30}}, \bibinfo{pages}{093123}
  (\bibinfo{year}{2020}).

\bibitem[{\citenamefont{Friedman and Landsberg}(2013)}]{doi:10.1063/1.4793782}
\bibinfo{author}{\bibfnamefont{E.~J.} \bibnamefont{Friedman}} \bibnamefont{and}
  \bibinfo{author}{\bibfnamefont{A.~S.} \bibnamefont{Landsberg}},
  \bibinfo{journal}{Chaos: An Interdisciplinary Journal of Nonlinear Science}
  \textbf{\bibinfo{volume}{23}}, \bibinfo{pages}{013135}
  (\bibinfo{year}{2013}).

\end{thebibliography}

\end{document}